\documentclass[twocolumn,prl,showpacs,multicol,amsmath,amssymb]{revtex4-1}
\usepackage{amssymb}
\usepackage{amsmath}
\usepackage{graphicx}
\usepackage{epstopdf}
\usepackage{bm}
\usepackage{gensymb}
\usepackage{soul}
\graphicspath{{Figs/}}

\newcommand{\be}{\begin{equation}}
\newcommand{\ee}{\end{equation}}
\newcommand{\bk}{{{\bf{k}}}}
\newcommand{\bq}{{{\bf{q}}}}

\newcommand{\bea}{\begin{eqnarray}}
\newcommand{\eea}{\end{eqnarray}}

\newcommand{\bd}{\begin{displaymath}}
\newcommand{\ed}{\end{displaymath}}
\newcommand{\ba}{\begin{array}}
\newcommand{\ea}{\end{array}}
\newcommand{\bi}{\begin{itemize}}
\newcommand{\ei}{\end{itemize}}
\newcommand{\bc}{\begin{center}}
\newcommand{\ec}{\end{center}}
\newcommand{\bfl}{\begin{flushleft}}
\newcommand{\efl}{\end{flushleft}}
\newcommand{\bfr}{\begin{flushright}}
\newcommand{\efr}{\end{flushright}}
\newcommand{\no}{\nonumber}

\newcommand{\Sr}{Sr$_2$IrO$_4$}
\newcommand{\mi}{\rm i}

\def\bk{{\bf k}} \def\bq{{\bf q}}  
 \def\hbg{\hat{{\bf g}}} \def\bd{{\bf d}}

\def\6{\partial}
  
   \def\e{\epsilon}

   \def\p{\pi}

\def\={\!\!\!&=&\!\!\!}
\def\+{\!\!\!&&\!\!\!+~}
\def\-{\!\!\!&&\!\!\!-~}

\usepackage{color}
\usepackage[dvipsnames]{xcolor}
\usepackage{xcolor}
\begin{document}
\title{Mixed-pairing superconductivity in $5d$ Mott insulators with   antisymmetric exchange} 
\author{Mohammad-Hossein Zare$^{1,2}$}\email{zare@qut.ac.ir}
\author{Mehdi  Biderang$^{3,4}$}
\author{Alireza Akbari$^{3,5,6}$}\email{alireza@apctp.org}
\affiliation{$^1$Department of Physics, Faculty of Science, Qom University of Technology, Qom 37181-46645, Iran}
\affiliation{$^2$School of Mathematics and Physics, University of Queensland, Brisbane, 4072 Queensland, Australia}
\affiliation{$^3$Asia Pacific Center for Theoretical Physics, Pohang, Gyeongbuk 790-784, Korea}
\affiliation{$^4$Department of Physics, University of Isfahan, Hezar Jerib, 81746-73441, Isfahan, Iran}
\affiliation{$^5$Department of Physics, POSTECH, Pohang, Gyeongbuk 790-784, Korea}
\affiliation{$^6$Max Planck POSTECH Center for Complex Phase Materials, POSTECH, Pohang 790-784, Korea}
%
%
\date{\today}
%
\begin{abstract}
 We investigate  the potential existence  of a superconducting phase in $5d$ Mott insulators with an eye to   hole doped Sr$_2$IrO$_4$.
Using a mean-field method,    a  mixed singlet-triplet superconductivity, $d + p$,    is observed due to the antisymmetric exchange   originating from    a quasi-spin-orbit-coupling.  
Our calculation on  ribbon geometry shows possible existence of the topologically protected edge states, because of nodal structure of the superconducting gap.
 These edge modes  are spin polarized and emerge as zero-energy flat bands,  supporting a symmetry protected Majorana states, verified  by evaluation of winding number and $\mathbb{Z}_2$ topological invariant.
At the end, a possible experimental approach for  observation  of these edge states  and determination of   the superconducting gap symmetry are  discussed based on  the quasi-particle interference (QPI) technique.
\end{abstract}
%
\maketitle
%
{\it Introduction:}
 Topological superconductors  are  one of the most outstanding topics  in condensed matter   because of  inevitably  hosting  Majorana fermions~\cite{Beenakker:2015aa}. 
In contrast to  artificial  
compounds   based on the proximity effect~\cite{Nadj-Perge:2014aa},  
 class  of  intrinsic  topological superconductor   may be realised  in heavy fermion noncentrosymmetric superconductors~\cite{Sato:2009aa}
   or transition metals oxides (TMOs)~\cite{Rau:2016aa}.
Strong correlated  interactions~\cite{Imada:1998aa,Brandow:1977aa} make  TMOs into a  veritable playground
for the study of novel exotic phases.
These  include   intriguing phenomena  such as high-temperature  superconductivity (SC) in cuprates,
colossal magnetoresistance, multi-ferroics, and different ordered magnetic phases~\cite{Tokura:2000aa,Dagotto:2005aa,Lee:2006aa,Scalapino:2012aa}.
This group of materials, especially the $3d$-TMOs~\cite{Zaanen:1985aa}, are  ideal systems for observing signatures of Mott insulating behaviour.
Recently, a new class of Mott insulators based on iridates,  
have attracted    immense   attentions~\cite{Witczak-Krempa:2014aa, Cava:1994aa,Shimura:1995aa,Moon:2008aa,Cao:2017aa}.
The belief that  their Mott phase  is motivated by the interplay of electron-electron and spin-orbit coupling (SOC). In these materials, $5d$ orbitals
are partially filled, and because of  the extended nature of the $5d$ orbital, the electron-electron interaction is much smaller
than that of  $3d$ Mott insulators. 
However,  the former  shows  a much larger SOC due to the fact that the strength of the SOC is controlled by %
the fourth  power of  the atomic number.  
Therefore, the intermediate interaction strength together with relatively large SOC makes them   a
unique  and subtle system  to study both theoretically and  experimentally. 

Among the 5d TMOs, the Iridates and specially  \Sr~is widely investigated for superconducting properties. 
This follows  from the  structural and electronic  similarities with  La$_{2}$CuO$_4$ and Sr$_{2}$RuO$_4$~\cite{Kim:2012ab,Fujiyama:2012aa,Cetin:2012aa}. 
In this context, predictions of SC  in   \Sr~arise from variational Monte Carlo simulations (VMC)~\cite{Watanabe:2013aa},  
the singular-mode functional renormalization group~ (SM-FRG)\cite{Yang:2014aa},  
 and dynamical mean field theory (DMFT)~\cite{Meng:2014aa}. %
However, the nature  of the achievable SC  is still under  debate.
While VMC proposes a $d$-wave superconducting phase only for electron doping, SM-FRG addresses two possible scenarios: a mixed singlet-triplet 
SC with $d^*_{x^2-y^2}$ symmetry for electron doped and a mixed singlet-triplet SC with $s^*_\pm$ symmetry for the hole doped cases.
In addition to these methods,  DMFT foresees a  topological $p + {\mi} p$ pseudo-spin (singlet $d$-wave)  pairing for hole (electron)~doped samples.
From experimental point of view,   the  existence  of Fermi arcs 
suggests   electron doped  
as  a potential candidate for the $d$-wave SC but with $T_c$  
 lower than that of La$_{2}$CuO$_4$~\cite{Kim:2014aa,Kim:2016aa}. 
 Furthermore, a signature of high Tc is found in electron doping~\cite{Yan:2015aa} and a  $p$-wave pairing state  is  reported  by substitution of Ru (hole-doping)~\cite{Yuan:2015aa}.

In this paper, by applying a  mean-field (MF) approach, we sketch  the symmetry and structure of the anomalous order parameter (OP) in  the iridates  
alluded to above.
We mainly focus on the broken inversion symmetry, 
 which  gives rise to a mixed singlet-triplet superconducting state to arise. 
This broken inversion symmetry, occurring as a result of   the bond-deviations,    introduces  the antisymmetric exchange   and   
   a quasi-SOC  (spin-dependent hopping).
    In particular, we deal with interesting  questions such as at what level of the doping or at what strength of the  quasi-SOC, 
 SC is recognized in the system, what is the symmetry of the superconducting OP,  and more importantly    is there a possibility  to find a topological SC 
 by calculating  the global  topological invariants.
We believe that our straightforward  MF slave boson approach answers these broad questions and opens doors for future analytical and numerical studies.
\\

{\it Model and Method:} 
The layered  \Sr~consists of two-dimensional  IrO$_2$ layers, in which Ir$^{4+}$ ions form a square lattice. 
 In contrast La$_2$CuO$_4$,
the IrO$_{6}$ octahedron in  \Sr~is elongated along
the $c$-axis and is rotated around it by an angle $\theta\approx11^{\circ}$\cite{Crawford:1994aa}. This deviation angle increases by decreasing of temperature~\cite{Chikara:2010aa}. 
On top of this, experimental techniques such as X-ray scattering and neutron diffraction indicate a canted antiferromagnetic order in this material~\cite{Moon:2009aa,Ye:2013aa}.
Moreover, the magnetic susceptibility measurement reveals a weak ferromagnetism in  \Sr~\cite{Cao:1998aa}.
Due to the strong SOC,
the spin and orbital degrees of freedom locally entangle and
$t_{2g}$ orbital splits further into fully filled  ${J}_{\mathrm{eff}}=3/2$ (lower), and half-filled    ${J}_{\mathrm{eff}}=1/2$ (upper) states. 
 This observation is a result of the reduced energy bandwidth 
arising from the presence of SOC, which in turn leads to a Mott insulating phase at even smaller $U$.
The theoretical studies~\cite{Watanabe:2010aa,Jackeli:2009aa} and several experiments~\cite{Kim:2012ab,Fujiyama:2012aa} support 
the single orbital picture of the ${J}_{\mathrm{eff}}=1/2$ Mott insulating phase in governing the low energy physics of Ir-oxides~\cite{Kim:2012aa}.
Based on this picture,
the single orbital  Hubbard 
model
is given by~\citep{Wang:2011aa,Jin:2009aa,Rau:2016aa}
%
\bea
\begin{aligned}
\hspace{-0.6cm}
{\cal H}
\!=
&
-
\!\!\!\!
\sum_{\langle ij\rangle,\alpha}
\!\!
t_0(\theta)
\;
c_{i\alpha}^\dagger c_{j\alpha}^{}
- 
\sum_{\langle ij\rangle,\alpha \beta}
\!\!
{\mi}
 (-1)^i
 \;
 t'_{0} (\theta)
 \;
 c_{i\alpha}^\dagger 
 \sigma_{\alpha\beta}^z 
 c_{j\beta}^{}
 \\
 &+
  \sum_{i}
 \! 
 U
 n_{i\uparrow}n_{i\downarrow}.
 \end{aligned}
\label{eq:H}
\eea
%
Where $t_0(\theta)= 2t_{0}/3\cos\theta (2\cos^{4}\theta - 1)$, and  $t'_0(\theta)= 2t_{0}/3\sin\theta (2\sin^{4}\theta - 1)$ are the hopping integrals~\cite{Jin:2009aa},
 and   $\boldsymbol{\sigma}$ is the vector of Pauli matrices acting on pseudo-spin space, 
 $c^{\dagger}_{i\alpha}~(c^{}_{i\alpha})$ stands for the creation
(annihilation) operator of an electron 
with  pseudo-spin $ \alpha=\uparrow,\downarrow$ on site $i$, 
and $ n_{i\alpha}=c_{i\alpha}^\dagger c_{i\alpha}^{}$ is a number operator.
 Here $U$ is   on-site Coulomb interaction, and the first two terms represent effective hopping integrals.
The second term is a quasi-SOC and its  spin-dependency  is formulated  from bonding deviation angle, $\theta$,
 from $180^{\degree}$ along the Ir-O-Ir bond angle,
 and behaves in a similar way as an intrinsic SOC.  
This bond angle can be controlled by tuning the chemical potential, $\mu$, or
 by applying a magnetic field~\cite{Ge:2011aa} or a strain~\cite{Kim:2017aa,Nichols:2013aa}. 
%
%
 \begin{figure}[t]
 \begin{center}
\vspace{0.10cm}
 \hspace{-0.1cm}
 \includegraphics[width=0.46 \linewidth]{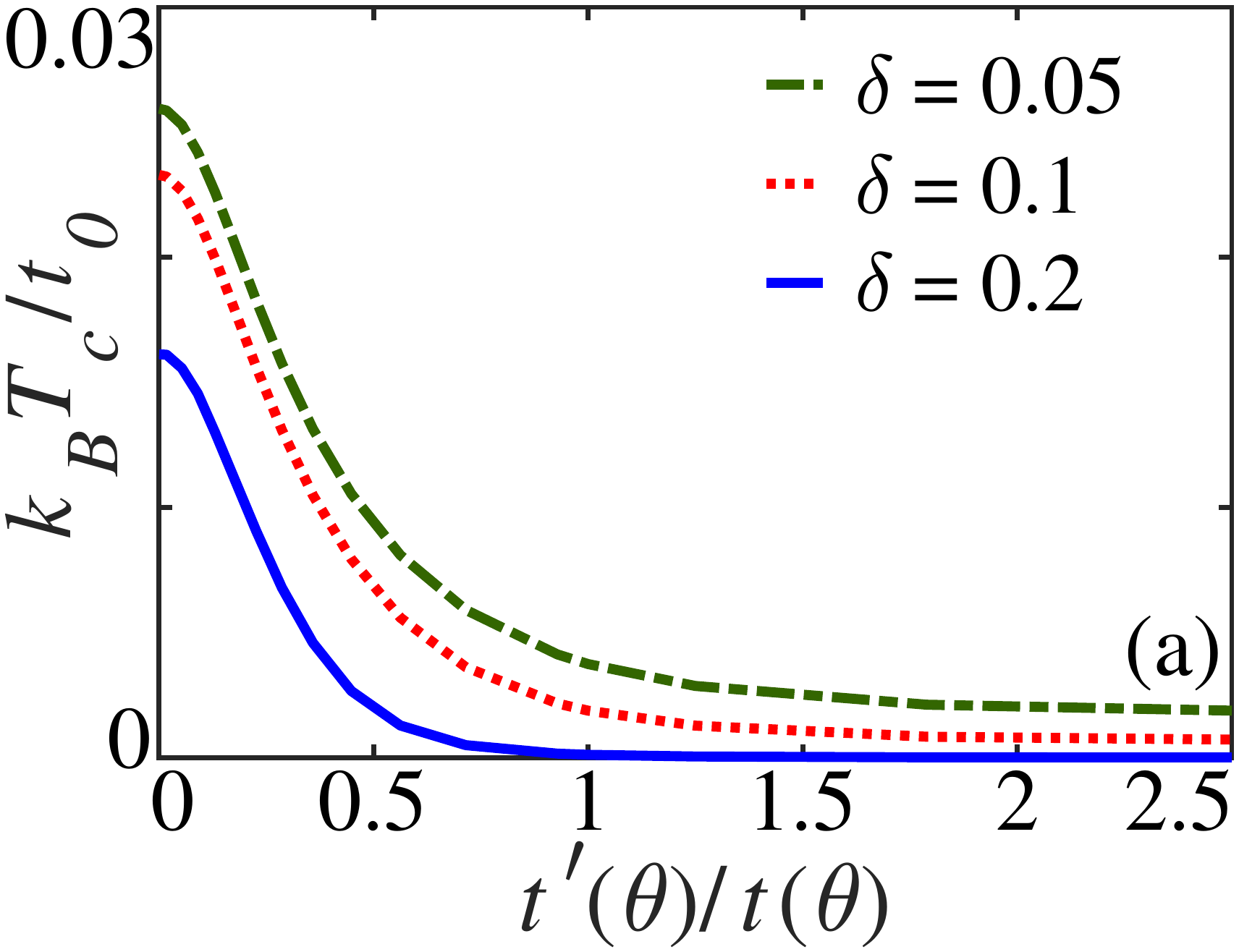} 
 \hspace{.0cm}
\includegraphics[width=0.518\linewidth]{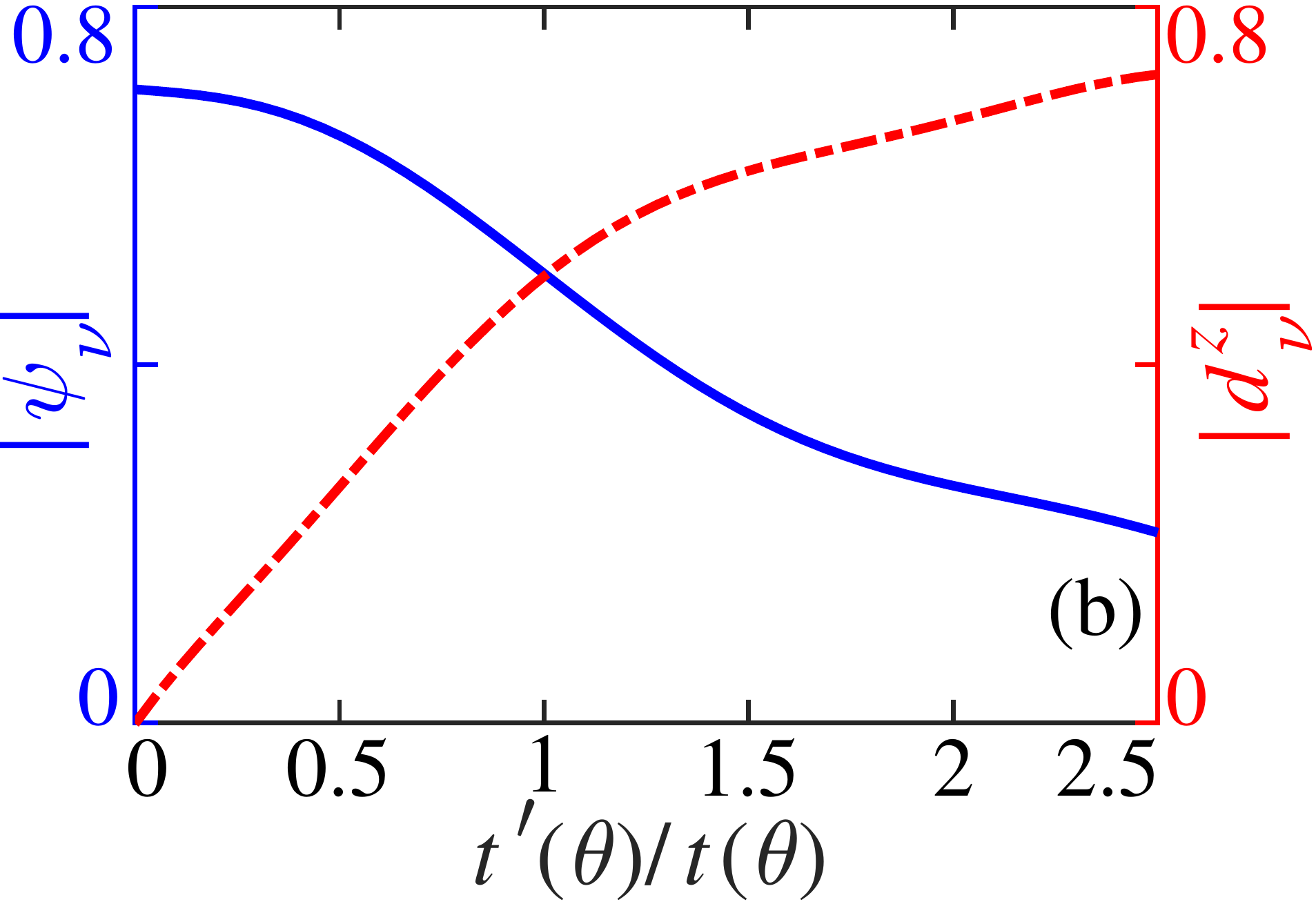}
%
\end{center}
 \vspace{-0.7cm}
\caption{(Color online) 
 (a) The transition temperature, $T_{c}$,  
 in the mixed singlet-triplet channel versus   quasi-SOC (spin-dependent hopping)  
 strength
${t' }$ for different doping levels, $\delta$. 
(b)  Magnitude of  MF order parameters, $\psi_{\nu}$  (blue-solid) and $d^{z}_{\nu}$ (dashed-red),
plotted as a function $t'$, for $\delta=0.1$.
}
\label{fig:1}
\end{figure}
%
%
%
%
At half-filling  the effective %
 Hamiltonian may be considered  as follows~\cite{Jackeli:2009aa,Ge:2011aa}, 
\begin{align}
{\cal H}_{J}=\sum_{\langle ij\rangle}
\Big[
J_H 
({\bf S}_{i}\cdot {\bf S}_{j}
-\frac{n_{i}n_j}{4}
)
+J_{z}S_{i}^{z}S_{j}^{z} + {\bf D}\cdot {{\bf S}_{i}}\times {\bf S}_{j}
\Big],
\label{eq:SH}
\end{align}
with 
antiferromagnetic 
Heisenberg exchange interaction $J_H$,  the Ising-like exchange interaction  $J_{z} $, as well as the antisymmetric Dzyaloshinskii-Moriya (DM)
exchange interaction perpendicular to IrO$_2$ layers.
The latter two terms in the effective exchange Hamiltonian  are originated from the  quasi-SOC %
term.
Here 
$J_{H}=
4[t_{0}(\theta)^2-t'_{0}(\theta)^{2}
]/U$, 
$ J_{z}=8{{t'_{0}}(\theta)^{2}}/{U}$, 
and
${\bf D}=-D{\hat{\bf z}}$
with
$D=
8t_{0}(\theta)
t'_{0}(\theta)
/U$.
Doping introduces itinerant fermionic feature of the system, and in a similar way as cuprates the possible SC can be appeared.
To study the superconducting transition temperature, $ T_c$, and also the symmetry of the anomalous pairing function, we start
from the 
$t$-$t'$-$J$ model as ${\cal H}_{t {t' } J}={\cal H}_{t}+{\cal H}_{t'}+{\cal H}_{J}$. 
Then, we rewrite the pseudo-spin 
operators in ${\cal H}_{J}$ as a bilinear of two fermion operators,
${\bf S}_{i}=\frac{1}{2} f^\dag_{i,\alpha}{\boldsymbol {\sigma}}_{\alpha\beta}f_{i\beta}$,
where  $f^{\dag}_{i\alpha}$ is the fermionic spinon creation operator~\cite{Wen:2002aa}.
Substituting  the pseudo-spin operators  
by the fermionic  
operators, and  
rephrasing the result in terms of spin singlet and triplet operators:
 $s_{ij}= (f_{i\uparrow }f_{j\downarrow }-f_{i\downarrow }f_{j\uparrow })/\sqrt{2}$,
 $t_{ij,x}=(f_{i\downarrow }f_{j\downarrow }-f_{i\uparrow }f_{j\uparrow })/\sqrt{2}$
 ,
$t_{ij,y}={\mi} (f_{i\uparrow }f_{j\uparrow }+f_{i\downarrow }f_{j\downarrow })/\sqrt{2}$,
$t_{ij,z}={\mi}(f_{i\uparrow }f_{j\downarrow }+f_{i\downarrow }f_{j\uparrow })/\sqrt{2}$,  the effective  Hamiltonian can be  reconstructed   as 
%
\bea
\begin{aligned}
\hspace{-0.73cm}
{\cal H}_{J}
\!
=
\!
&
-
\!
\sum_{\langle ij\rangle}
\Big[
J_H s^\dag_{ij}s_{ij}
+\frac{D}{2}(s^\dag_{ij}t_{ij,z}+t^\dag_{ij,z}s_{ij}^{})
 \\
 &
+\frac{J_z}{4}(s^\dag_{ij}s_{ij}-t^\dag_{ij,x}t_{ij,x}-t^\dag_{ij,y}t_{ij,y}^{}+t^\dag_{ij,z}t_{ij,z}^{})
\Big].
\end{aligned}
\label{eq:Heff}
\eea
%
Therefore, the  MF Hamiltonian can be achieved
by adopting   the spin-singlet and spin-triplet  MF OPs: $\psi_{\nu}=  \langle s_{ij}\rangle/\sqrt{2}$  and $d^{\gamma}_{\nu}= \langle t_{ij,\gamma} \rangle/ \sqrt{2}$.
In which, ${\nu\in\{x,y\}}$ indicates the direction of the bonds on the square lattice,  and  $\gamma$ characterizes the component of the pairing triplet vector. Note that 
we use  the Kotliar-Ruckenstein 
 slave-boson formalism~\cite{Kotliar:1986aa} to ensure the Gutzwiller projection.
In this formalism,  
the original electrons are replaced with
four  auxiliary bosons and one spinful fermion,  that  at  half filling
 the doublon fields cannot be created at any doping levels.
Thus, in  MF hopping integrals are renormalized by factor of $\eta=2\delta/(1+\delta)$ that $\delta$ determines the hole doping values, and shows itself in the Hamiltonian by marking 
 $t=t(\theta)=\eta t_{0}(\theta)$,~and~$t'=t'(\theta)=\eta t'_{0}(\theta)$. 
Finally, by Fourier transformation (FT) of the fermionic  operators,
the total MF Hamiltonian  
can be written as,
%
\begin{align}
{\cal H}_{M\!F}
\!=
\!\!
\sum_{{\bf k}\alpha\beta } {
(
\varepsilon_{\bf k}
\!
+
\!
\bf{g_k}
\!
\cdot 
\!
\boldsymbol{\sigma}
)_{\alpha\beta}} 
f^{\dagger}_{{\bf k}\alpha}f_{{\bf k}\beta}^{} 
+
(\Delta^{\alpha\beta}_{\bf k} f^{\dagger}_{{\bf k}\alpha}f^{\dagger}_{-{\bf k}\beta}
\!
+
\!
h.c. ),
\label{eq:Hsc}
\end{align}
%
%
where the dispersion is defined by 
 $\varepsilon_{\bf k}=-2  t(\theta)
 (\cos k_x+\cos k_y)-\mu$,
 and 
 the Rashba type term,
 ${\bf{g_k}}=-2
 t'(\theta)
 (\sin k_x+\sin k_y){\hat{\bf z}}$, is originated from the quasi-SOC. 
 Because of a rotation of the oxygen octahedra,  the inversion symmetry is  locally broken and a DM interaction allows for the non-cubic structures
 like Sr$_2$IrO$_4$.
  As a result, the spin singlet and triplet pairings coincide in the  gap function,  i.e.,
$
\Delta_{\bf k}= {\mi}
(\psi_{\bf k}\sigma_0+{\bf d}_{\bf k}\cdot {\boldsymbol \sigma}
)
\sigma_y
$,
and their contributions  are accessed via 
$
\psi_{\bf k} =1/4\sum_{\nu} \big[ (4J_H+J_z) \cos k_{\nu} \psi_{\nu} +2D  \cos k_{\nu} d^z_{\nu}\big] ,
$
$ 
d^x_{\bf k}  = -{\mi}  J_z/4 \sum_{\nu} \sin k_{\nu} d^x_{\nu},
$
$ 
d^y_{\bf k} = {\mi} J_z/4 \sum_{\nu}  \sin k_{\nu}  d^y_{\nu} ,
$
 and 
$
d^z_{\bf k}  =1/4 \sum_{\nu} \big[ 2D \sin k_{\nu} \psi_{\nu} + J_z \sin k_{\nu} d^z_{\nu}  \big].
$
Close to
$ T_c$
the MF parameters, $\psi_{\nu}$ and $d^{\gamma}_{\nu}$, are suppressed.
In order to find  $ T_c$, it is necessary to obtain the eigenvalues of stability matrices for the spin-singlet and spin-triplet OPs,
where the critical temperature is determined by finding the largest temperature that at least one of the eigenvalues of different channels %
be one~\cite{Black-Schaffer:2007aa,Hyart:2012aa}.
%

%
 \begin{figure}[t]
 \begin{center}
 \hspace{-0.4cm}
 \includegraphics[width=0.47  \linewidth]{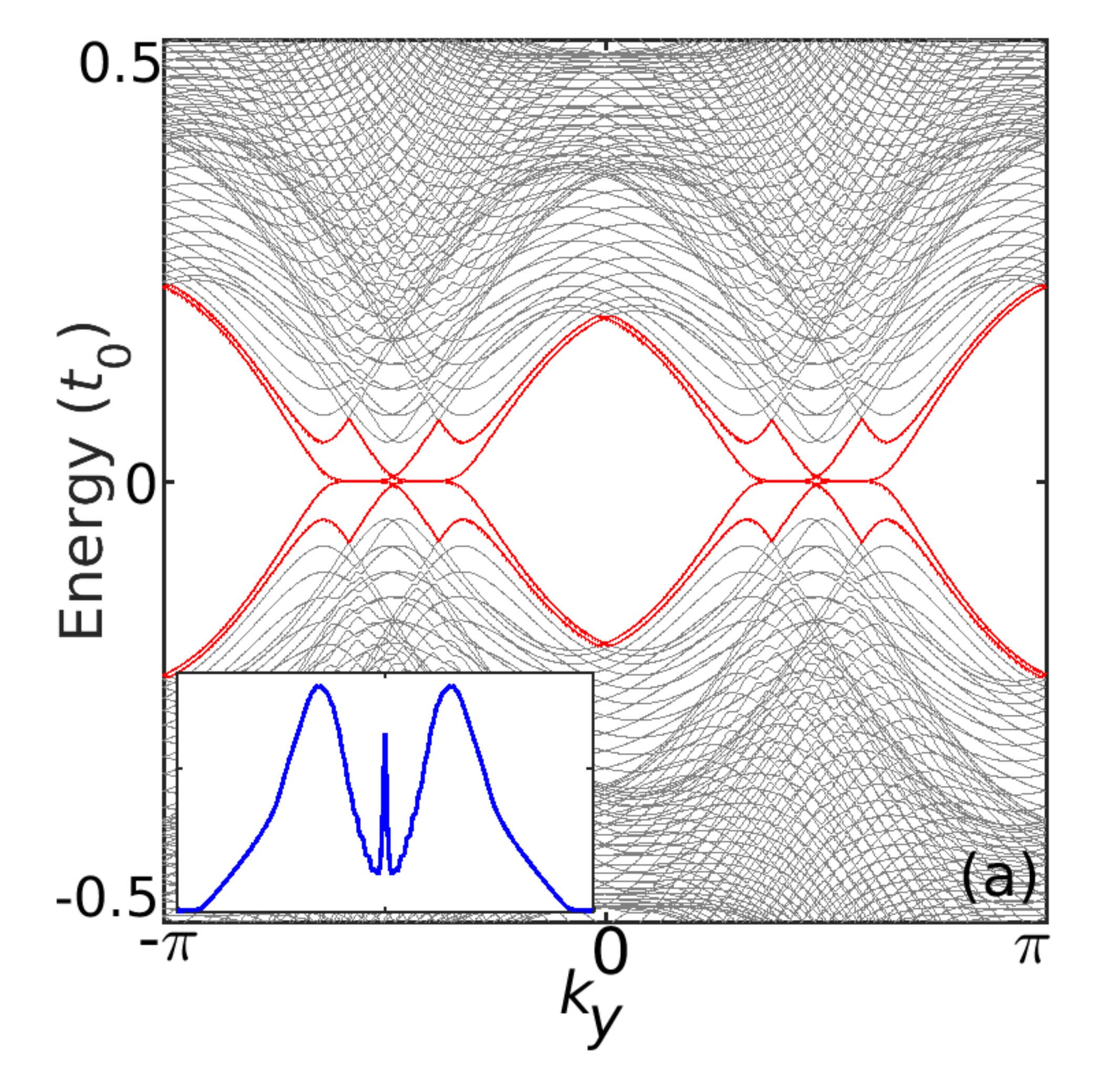}
  \hspace{-0.32cm}
 \includegraphics[width=0.58 \linewidth]{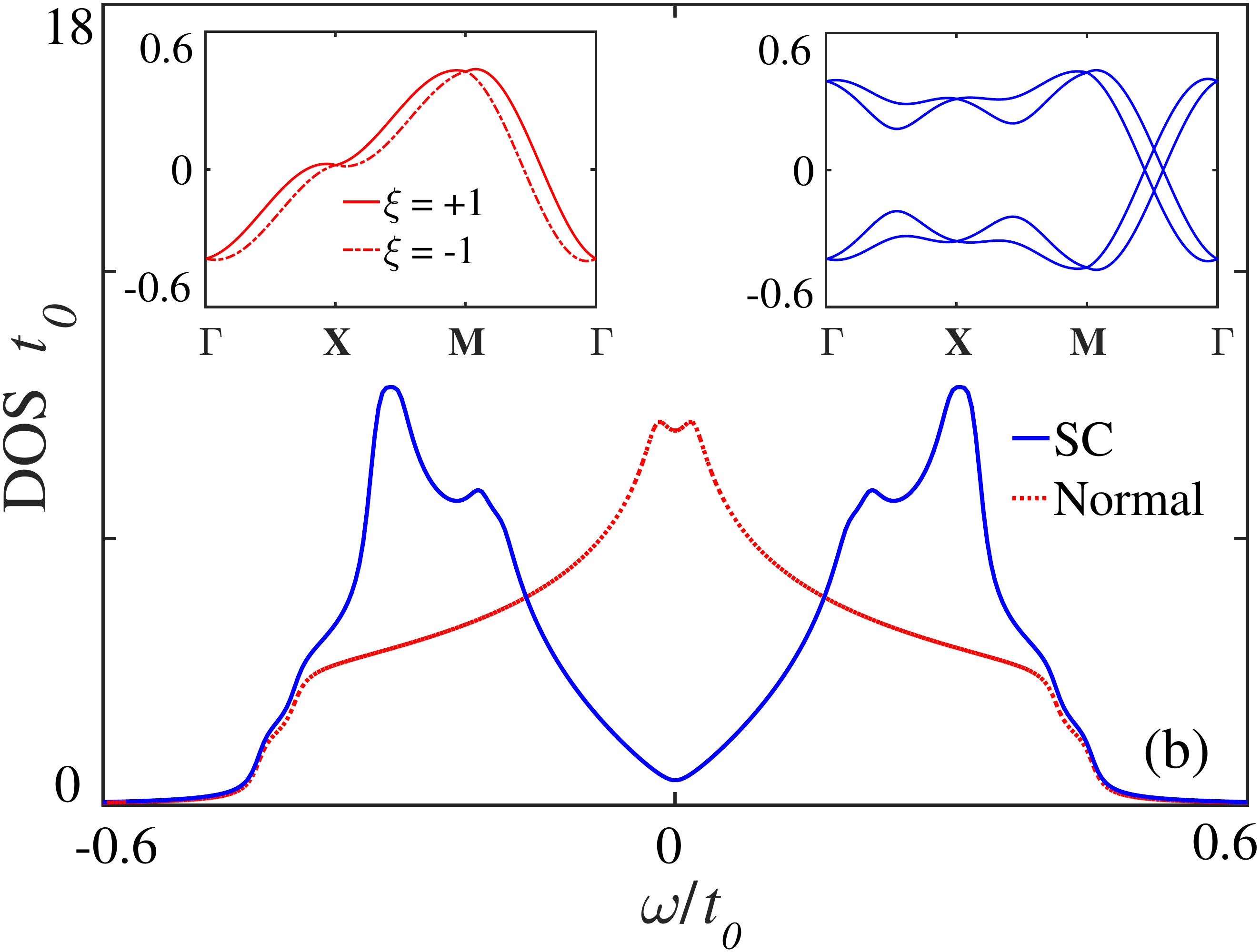}   
  \\
  \vspace{-0.1cm}
   \hspace{-0.53cm}
  \includegraphics[angle=0,width=1.05 \linewidth]{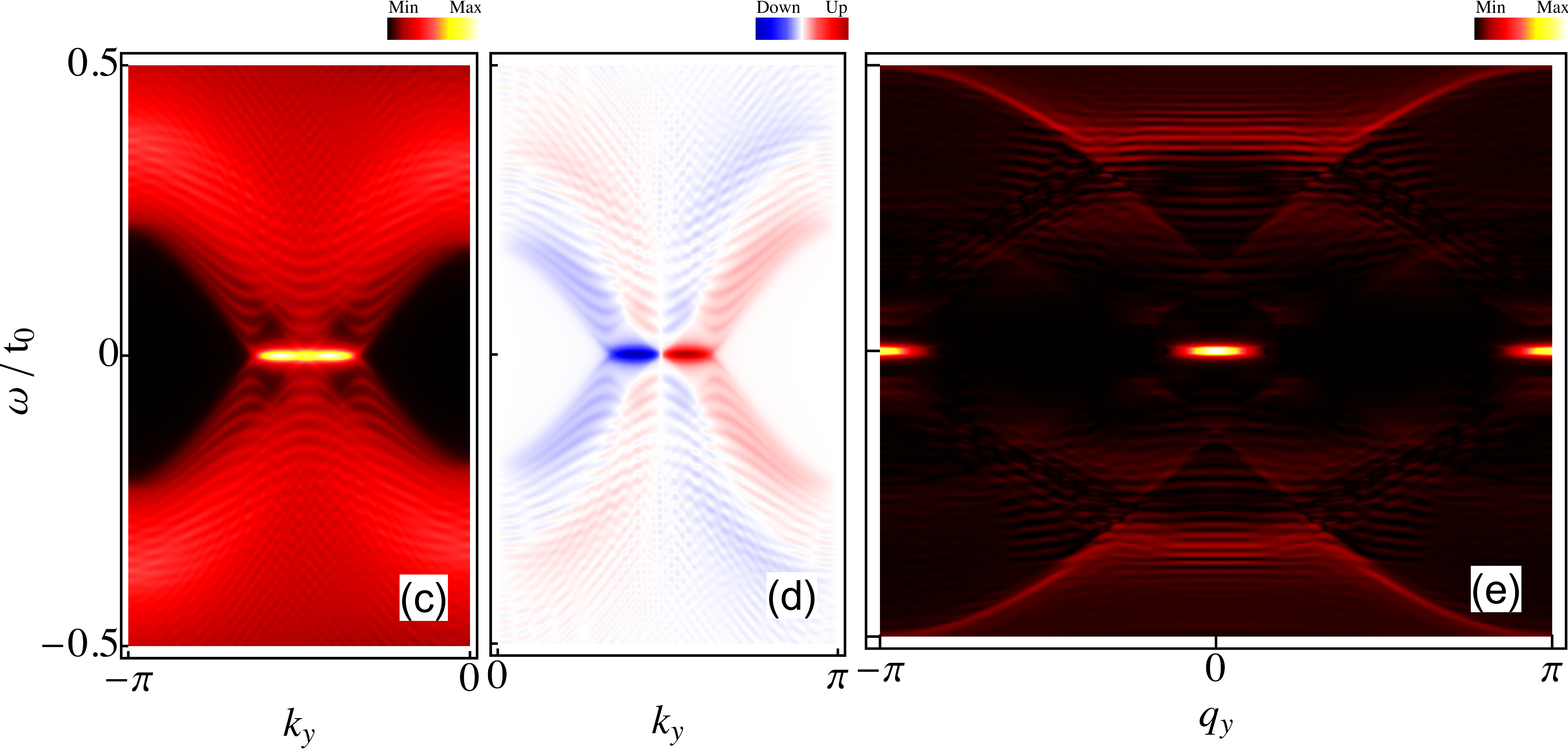}
\end{center}
 \vspace{-.75cm}
\caption{(Color online) 
(a)   
 The Bogoliubov electronic dispersion 
 of hole doped \Sr~on a ribbon for the topological phase; the red bands indicate the topological in-gap  edge states with 
  two-fold degenerate zero-energy Majorana-type with flat  bands. 
  Inset represents the edge LDOS of the superconducting state, where the zero energy pick is originated from the   zero-energy flat  bands, the range of the plot is the same as (b).
(b) Bulk  LDOS in normal (red-dashed) and superconducting  (blue-solid) phases.
Insets: left shows the band structure of quasi-spin-orbit split bands: $\xi=\pm 1$, in normal phase,
and right presents the quasi-particle bands in   superconducting phase.
(c,d) 
The intensity plots of  the momentum- and spin-resolved LDOS
along the   $\Gamma$-Y momentum direction, respectively. The up and down helical states in (d)  correspond to the states with  the winding numbers $+1$ and $-1$, respectively. 
(e)  Intensity of the QPI dispersion  (absolute value) at the edge (slab: $n= 1$) for the Y$'$-$\Gamma$-Y momentum direction.
 }
\label{fig:2}
\end{figure}
%

In presence of the DM term,  the spin-singlet 
and   $z$-component of  spin-triplet OPs
are coupled, while  in-plane terms  
remain untouched. Therefore, the latter terms, $d^{x}_{\nu}$ and $d^{y}_{\nu}$, have similar stability matrices as well as the  set of eigenvalues with two-fold degeneracies.
Comparing the critical temperature  
of  the mixed singlet-triplet  and the in-plane triplet channels,
supports an admixture state of a $d$-wave state $d_{x^2-y^2}$ 
 (nodal) and a $p$-wave
state $p^{z}_{x-y}$. 
This is originated from the fact that the spin-triplet component should be aligned to   quasi-SOC  direction,  ${\bf{d_k}}\parallel  {\bf{g_k}}$, by its similarity to 
 pure   intrinsic SOC~\cite{Frigeri:2004aa}.
It should be noted that this admixture state is invariant under the time-reversal symmetry. 
 Fig.~\ref{fig:1}(a) shows the critical temperature of the  state as a function of $t'$ for different
doping levels, we set  the Coulomb interaction $U\sim 6t_0$~\cite{Watanabe:2013aa}.
To investigate the importance of  spin-dependent hopping term in creating  the admixture state,  the magnitude of the MF parameters, $\psi_{\nu}$ and $d^{z}_{\nu}$, versus  
 $t' $ are shown in Fig.~\ref{fig:1}(b).
They represent that at  $t' = 0$, OP is pure singlet, however its triplet part   starts  to grow by increasing of the quasi-SOC, %
while
 simultaneously the singlet  part begins to decrease.
It should be emphasized  that the existence of the mixed superconducting
phase  
can support Majorana edge-modes by closing  the bulk  SC gap.  
These topological  edge-modes can be characterized  via calculating the momentum dependent winding number for the edge states, which shows  changes as a result of projection of the bulk-gap nodes~\cite{Sato:2011aa,Tanaka:2012aa,Schnyder:2013aa}. 
 \\
{\it Topological invariants: }
To study the edge modes of superconducting state, we consider the system as a ribbon with open and periodic boundary conditions along $x$- and $y$-directions,
respectively. 
Employing  the 
``Altland and Zirnbauer ten-fold"
classification~\cite{Schnyder:2008aa,Chiu:2016aa}  
puts the Hamiltonian (Eq.~\ref{eq:Hsc}) in the
``class DIII",
with the ``global $\mathbb{Z}_2=1$ topological number". 
The energy spectrum of the admixture SC,  versus the momentum $k_y$ for  
$\theta=11^{\degree}$  %
is shown in Fig.~\ref{fig:2}(a), while the red lines indicate the degenerate edge states. 
Here and in the rest of the paper, we fix   $\delta=0.1$ ($\mu=-0.02t_0$) corresponding to deviation angle $\theta=11^{\degree}$, thus 
$\psi_{x}=-\psi_{y}=0.7$,
$d_{x}^z=-d_{y}^z=-0.12$, and since the mixed superconducting phase is stable  the remaining MF parameters are zero. 
In  the insets of Fig.~\ref{fig:2}(b), we show the  band structure of quasi-spin-orbit split bands: $\xi=\pm 1$, in normal phase (left panel),
and   the quasi-particle bands in   superconducting phase (right panel).
The bulk  density of states (LDOS) in normal (dashed line) and superconducting  (solid line) phases are presented in  Fig.~\ref{fig:2}(b).
To investigate the topological nature of SC  phase, we calculate the   momentum- and spin-resolved LDOS in Fig.~\ref{fig:2}(c\&d), respectively. In particular, the polarization of spin-resolved LDOS shows strong momentum dependency and its sign tendency  at each flat bands can be interpreted  as  changing of winding number, $W=\pm 1$, for corresponding  bands.
As it is shown in the right inset of Fig.~\ref{fig:2}(a), these protected states lead to finite density of states at zero energy, which can be detected as a zero bias hump in the $dI/dV$ 
curve of scanning tunneling microscopy (STM) at the edge of the sample. 
The general form of the finding results are qualitative the  same by changing the deviation angle, before the SC becomes unstable for the larger angles.
\\

{\it Quasiparticle interference (QPI): }
The STM-based QPI  is one of the remarkable techniques to picture the possible existence of a superconducting state and  investigating  its pairing symmetry.
 It can show  information on the gap symmetry in connection with its dependency on the phase of the superconducting OPs and also on the form and change of the constant energy contours~\cite{Singh:2008aa,  *Knolle:2010aa,*Hanaguri:2010aa,*Allan:2012aa,*Akbari:2011aa, *Allan:2013aa,*Zhou:2013aa}.
  Here, we  provide an innovative way to probe the zero-energy nontrivial modes using QPI. 
  For this purpose, the system is divided  into parallel 
   lines (n=$1,2,\cdots,N$) along $y$-direction, which contains the nodes of the superconducting gap. Therefore, the Bogoliubov-de~Gennes (BdG) Hamiltonian is expressed as 
  ${\cal H}_{BdG}=\frac{1}{2}\sum_{k_y}\Psi^{\dagger}_{k_y}{\cal H}_{k_y}\Psi_{k_y}$, in which $\Psi_{k_y}^{\dagger}=(\Phi_{1,k_y}^{\dagger},\Phi_{2,k_y}^{\dagger},\cdots,\Phi_{N,k_y}^{\dagger})$
   is the generalized 
Nambu basis with $\Phi_{n,k_y}^{\dagger}=(f_{n,k_y \uparrow}^{\dagger}, f_{n,k_y \downarrow}^{\dagger}, f_{n,-k_y \uparrow}, f_{n,-k_y \downarrow})$. 
Then, the recursive relation for describing the superconducting slabs in the basis of Nambu spinors,  $4\times4$ matrix representation, is followed by
$$ T\Phi_{n-1,k_y}+M\Phi_{n,k_y}+T^\dagger\Phi_{n+1,k_y}=\zeta \Phi_{n,k_y},$$
 with the boundary condition: $\Phi_{0,k_y}=\Phi_{N+1,k_y}=0$. Here,  $\zeta$ represents the eigenvalue matrix of the ${\cal H}_{BdG}$ whose eigenstates  decay exponentially along $x$.
 $T$ defines  the hopping matrix connecting the nearest-neighbor slabs,
 and
 $M$  encodes the intra-hopping between degrees of freedom inside each slab as well as  on-site energies: 
%
\begin{align}
\label{eq:Ham_Slab}
 M=
 &\no
 -(2t \cos k_y+\mu)\vartheta_{z0}
 +
 [
\psi_yD+\frac{d^z_y}{2}
]
 \sin k_y \vartheta_{xx}
  \\&\no
- [
\psi_y(2J+\frac{J_z}{2})+d^z_y
] \cos k_y
\vartheta_{yy}
 -2t' \cos k_y \vartheta_{0z},
  \\
T=
 &\no
- t \vartheta_{z0}
-
{\mi}
t' \vartheta_{0z}-(D+\frac{J_z}{2}) ({\mi}\vartheta_{xx}+\vartheta_{yy})
- 2J  \vartheta_{yy},
 \end{align}
here    
$\vartheta_{\alpha \beta}=\tau_\alpha \otimes \sigma_\beta$, 
and 
${\tau}_\alpha$
are Pauli's matrices acting  on particle-hole space. 
The practical way to determine QPI  is to find FT of  spatial modulation of STM data due to 
the elastic scattering of quasiparticles
 from  impurity potential, $V$, which in the full Born approximation  is proportional to  changes in local density of states, 
  given~by
 %
%
\begin{equation}
\begin{aligned}
\label{eq:Ham_Slab}
\delta N(q_y,\omega)=-\frac{1}{\pi} V
{\rm Im} 
\Big[ \Lambda(q_y,{\mi}\omega)
\Big]_{{\mi}\omega\rightarrow \omega+{\mi}0^{+}}.
 \end{aligned}
\end{equation}
%
Hence,  QPI intensity for the  $s^{\rm th}$ slab, using the 
  t-matrix formalism as outlined in Refs.~\cite{Akbari:2013ab, *Akbari:2013aa, Lambert:2017aa}, is obtained as
%
%
\begin{align}\nonumber
\label{eq:Ham_Slab}
    \hspace{-0.2cm}
\Lambda_s(q_y,{\mi}\omega)
\!\!
=\!\!
 \sum_{k_y}
 {\rm Tr}_{\sigma}  
 \Big[P_{\tau} \hat{\rho}\hat{G}(k_y,{\mi} \omega)\hat{\rho}\hat{G}^{^\intercal} 
 \!
 (k_y-q_y, {\mi}\omega) 
 \Big]_s
,
 \end{align}
%
where index $s$ indicates the tracing over the elements in individual  block matrix of the slab $s$. Here the projector $P_\tau=I\otimes (\vartheta_{00}+\vartheta_{z0} ) /2$ and interaction $\hat\rho=I\otimes \vartheta_{z0}$  
 act on the generalised Nambu space, in which $I$  is an identity  matrix of size $N$.
Moreover, the retarded Green's function $\hat{G}(k_y,{\mi}\omega)\!=\!({\mi}\omega\!-\!{\cal H}_{k_y})^{-1}$ is defined for full slab geometry. 
The results of the intensity plots for the spectral function dispersion and the corresponding QPI dispersion for  the edge  state (slab: $n=1$)  are shown in the Fig.~\ref{fig:2}(c-f).
Indeed the edge state modes can easily be seen as  zero-energy flat bands in both spectral functions (momentum and spin-resolved LDOS) in Fig.~\ref{fig:2}(c\&d), as well as, the  QPI dispersions in Fig.~\ref{fig:2}(e). These flat dispersion bands  are  absent for the bulk state (middle slab).
We note that, to detect these edge states experimentally, one should measure the QPI in [100]-plane, since there is no dispersion along the $k_z$ direction.  

%
 \begin{figure}[t]
\centering
{ \includegraphics[width=0.4 \linewidth]{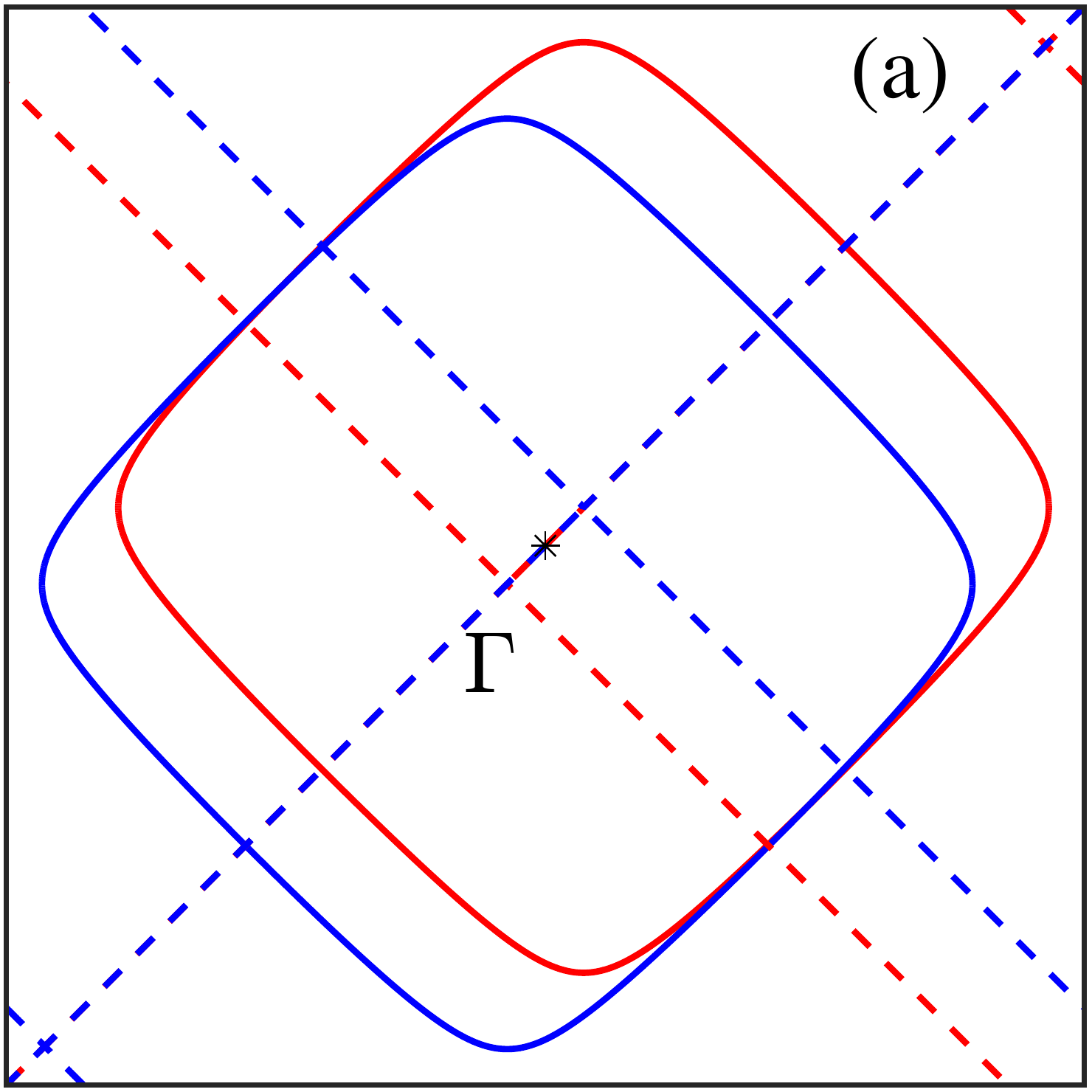}
 \hspace{0.0cm}
 \includegraphics[width=0.4 \linewidth]{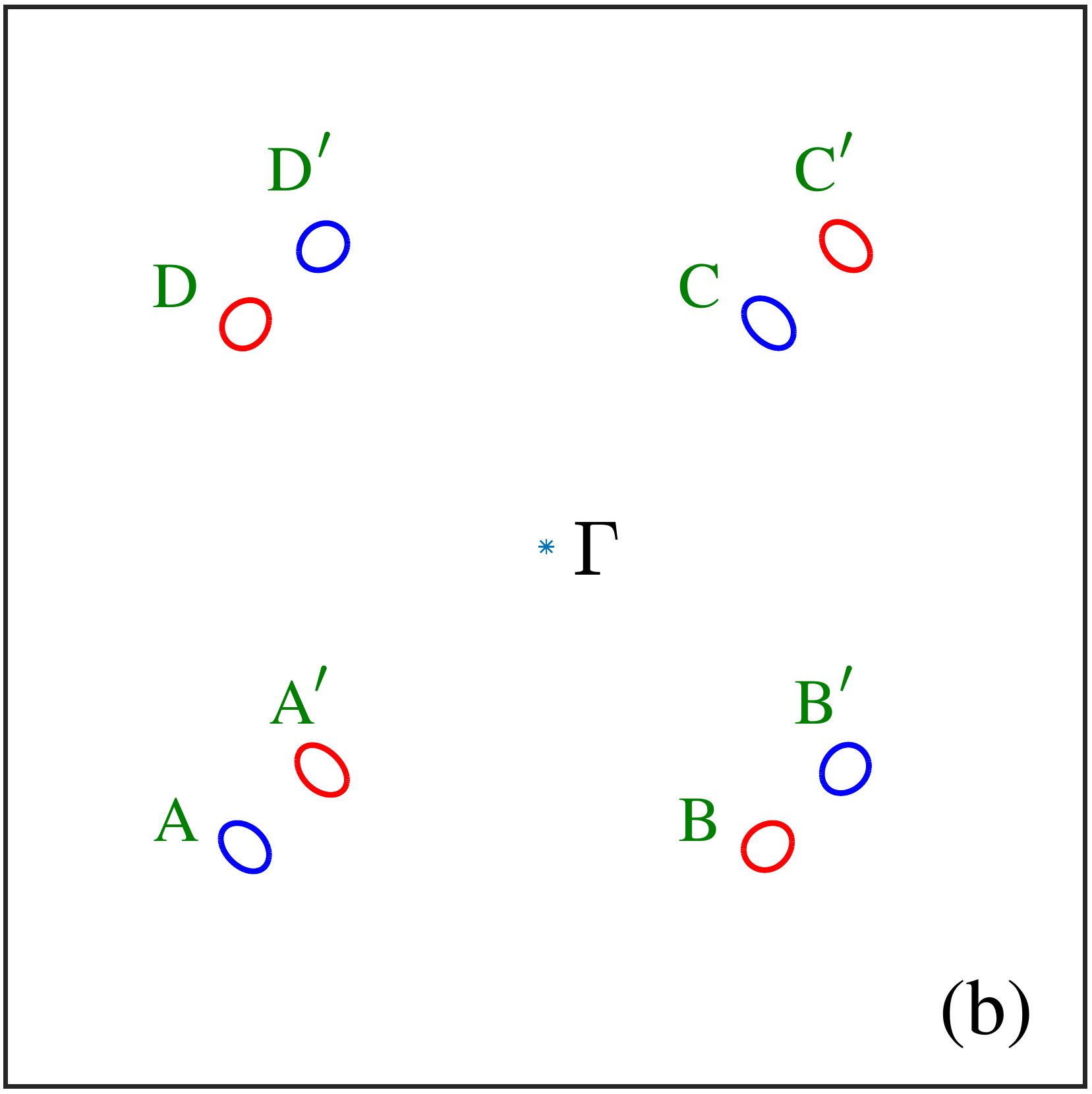} \\
\vspace{0.051cm}
\hspace{-0.0710cm}
 \includegraphics[width=0.4 \linewidth]{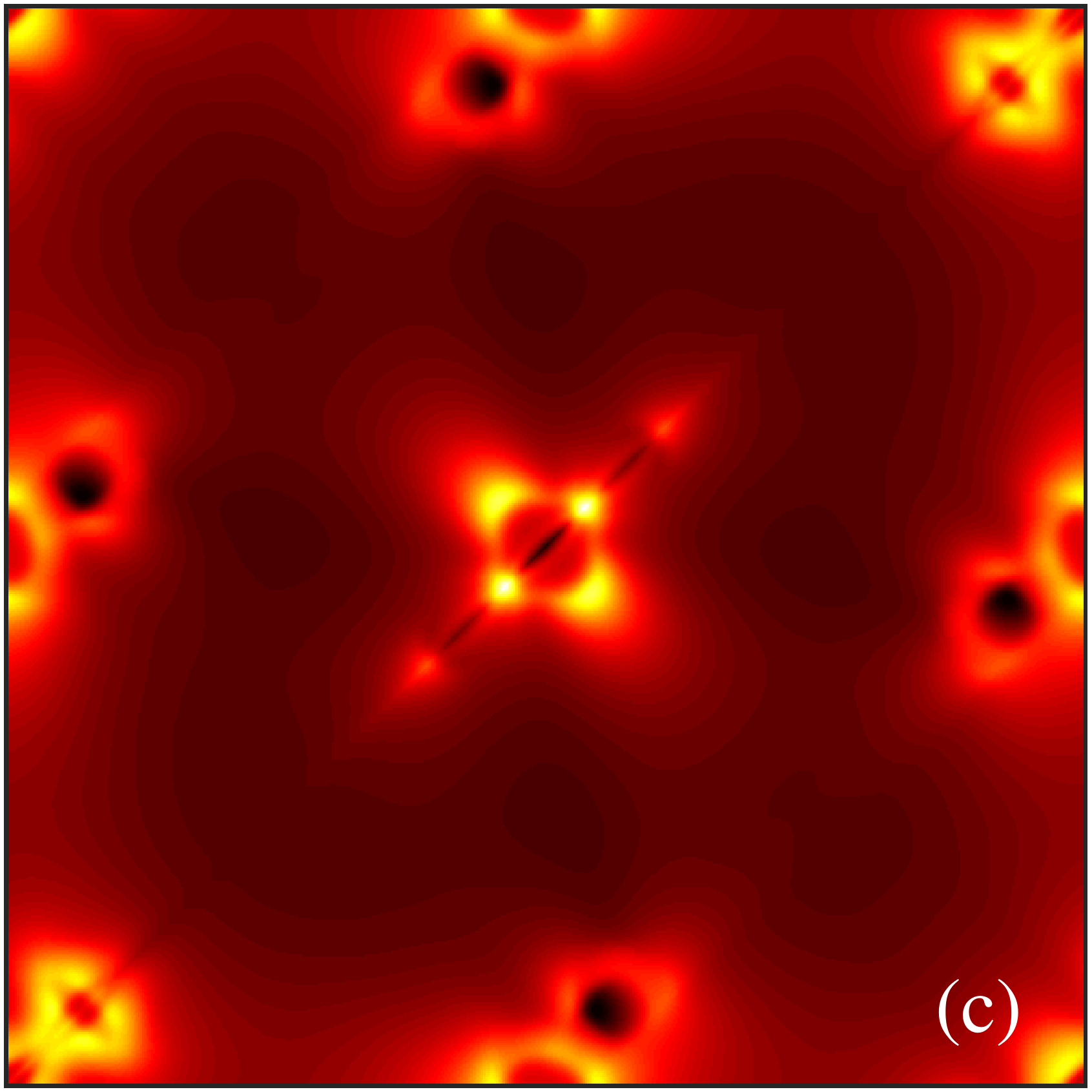}
\hspace{0.0cm}
  \includegraphics[width=0.395 \linewidth]{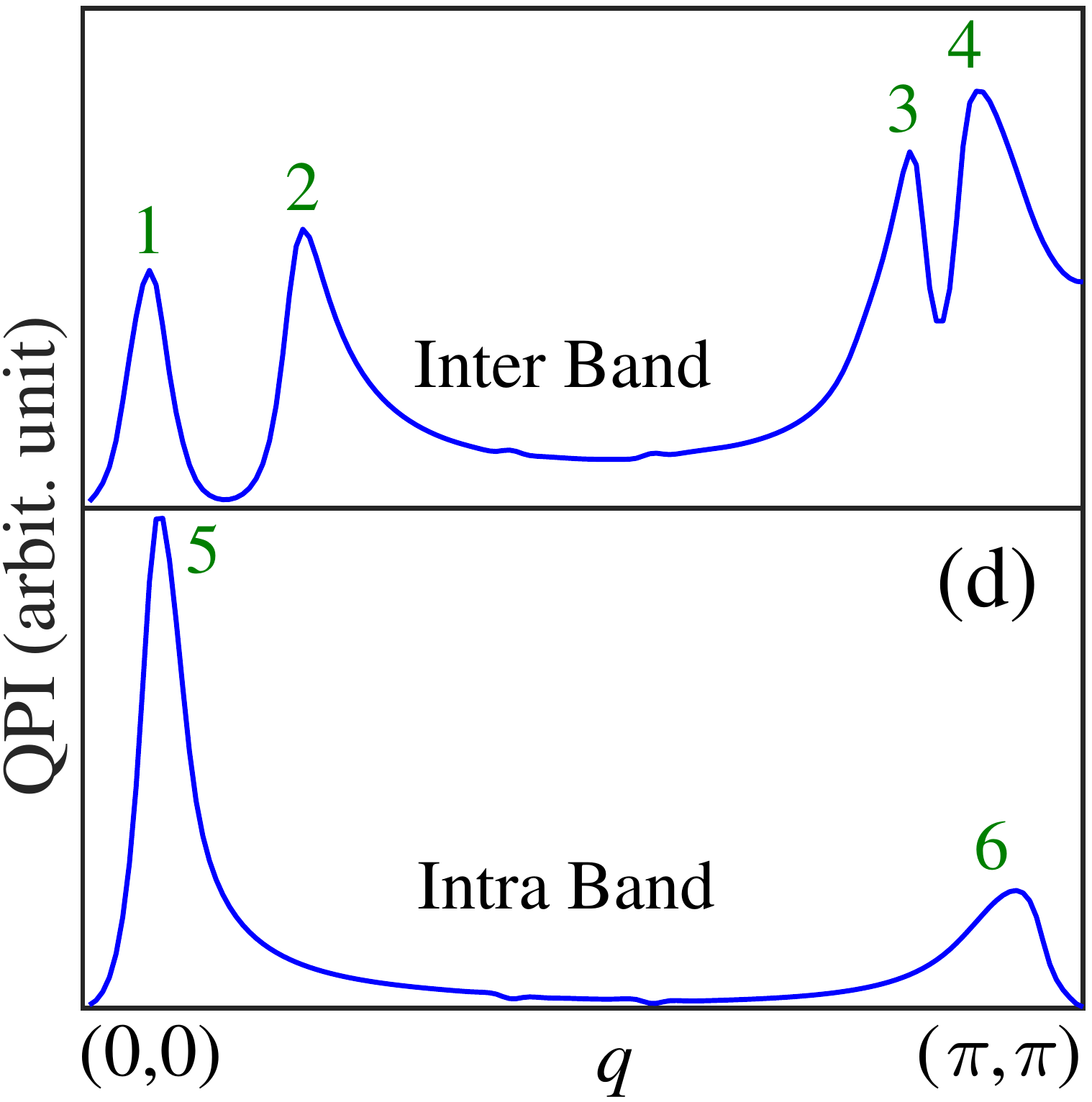}
\vspace{-0.24cm}
   }
\caption{(Color online) 
(a) Fermi surface energy contours (solid lines): blue (red) represents the band with $\xi=-1 \; (\xi=+1)$; the dashed lines indicate the node structure of the  gap function.    
(b) Constant  contours of energy (spectral function)   for the   superconducting state, in $({k}_{x},{k}_{y})$-plane  at 
$\omega=0.1t_0$. The small pockets, A, A$'$, $\ldots$, appear around the nodes in the spectral function, moderate  the scattering wave vectors. 
(c)
The  absolute value of the total QPI spectrum in $({q}_{x},{q}_{y})$-plane,  corresponding to spectral function in (b).
Note: plots range in  a-c  are $[-\p,\pi]$.
(d) The QPI intensity along the $(0,0) - (\pi,\pi)$ direction, for inter- (intra-) band scattering in upper (lower) panel. 
The peaks $(1-6)$ are related to  the main scattering vectors, namely
$1\&2$: BB$'$, DD$'$,
$3\&4$: A$'$C \& AC$'$, 
$5$:  BB \& B$'$B$'$ (DD \& D$'$D$'$),
and
$6$: AC \& A$'$C$'$. %
}
\label{fig:3}
\end{figure}
%

%
Finally,  to explore  the nodal structure of  the suggested %
 superconducting state,  the QPI spectrum  has been presented  in the $xy$-plane (bulk QPI), %
which~comes~by~\cite{Akbari:2013ab}
%
%
\be \nonumber
\begin{aligned}
\label{eq:Lambda}
\Lambda(\bq,{\mi} \omega_n)=
&\frac{1}{4N}\sum_{\bk\xi\xi'}
(1+\xi\xi' \hbg_\bk\cdot\hbg_{\bk-\bq})
\times
\\
&
\hspace{-1cm}
\frac
{({\mi} \omega_n+\e_{\bk\xi})({\mi} \omega_n+\e_{\bk-\bq\xi'})-\Delta_{\bk\xi}\Delta_{\bk-\bq\xi'}}
{[({\mi} \omega_n)^2-E^2_{\bk\xi}] [({\mi} \omega_n)^2-E^2_{\bk-\bq\xi'}]}
,
\end{aligned}
\ee
%
where the effective  quasiparticle energies,  
 $E_{\bk\xi}=\sqrt{\epsilon_{\bk\xi}^2+\Delta_{\bk\xi}^2}$,
 related to  the quasi-spin-orbit split-dispersions ($\xi=\pm1$)
defined by $\epsilon_{\bf k\xi}=\varepsilon_{\bf k}+\xi~|{\bf{g_k}}|$
 with superconducting gap
$\Delta_{{\bf k}\xi}=\psi_{\bf k}+\xi~d^{z}_{\bf k}$.
%
%
We present in the Fig.~\ref{fig:3}(a)  the normal state electron
 Fermi surface,  where the gap zeroes are displayed as dashed lines, and the resulting 
 split bands: $\xi=-1$ (blue), and  $\xi=+1$ (red)  
  are shown in the insets of Fig.~\ref{fig:2}(b).
The contours of the  quasiparticle energies, $E_{\bk\xi}$, at constant energy $\omega=0.1t_0$, are schemed  in  Fig.~\ref{fig:3}(b). The small pockets, appear around the nodes in the spectral function,  
 lead the scattering wave vectors  that play the main role in QPI pattern as  shown in  Fig.~\ref{fig:3}(c). 
 To better tracing of the scattering vectors, we show the QPI intensity along the  $(0,0) - (\pi,\pi)$ direction, for both inter- and intra- band scatterings process in  Fig.~\ref{fig:3}(d).
\\

{\it Summary:}
By focusing on   hole doped  \Sr,~we predict that the mixed singlet-triplet superconductivity   can  exist in  layered $5d$ transition metal oxides, as an example of  the new class Mott insulators.  
Our  results  demonstrates that the Dzyaloshinskii-Moriya interaction  plays an important role in finding  this interesting novel phase  
by preserving the time-reversal symmetry.
This also conjectures the existence of a mixed-paring phase, boosted by  antisymmetric exchange, in other iridates that host a similar  mechanism  
for an  insulating state.
This insulating  state  is confirmed for other iridates  such as Sr$_3$Ir$_2$O$_7$~\cite{Moretti-Sala:2015aa,Lu:2017aa} and BaIrO$_3$~\cite{Laguna-Marco:2010aa}. 
This analysis can be extended to models, which also
include the second-neighbour 
and the third-neighbour Heisenberg coupling.
Together with this prediction, we present a possible method for experimentally observing the zero-energy nontrivial edge states by STM-based QPI approach.  
\\
%

{\it Acknowledgments:}
We would like to thank G. Khaliullin, B.J. Kim, A. Schnyder, P. Thalmeier, P. Fulde, R. H. Mckenzie, A. R. Wright,  M. Kargarian, A.G. Moghaddam,  J. Yu, and E. Taghizadeh,   for  
useful discussions.
M.H.Z 
thanks Asia Pacific Center for Theoretical Physics (APCTP) for hospitality. 
A.A. acknowledges support through  NRF funded by MSIP of Korea (2015R1C1A1A01052411), and by  Max Planck POSTECH / KOREA Research Initiative (No. 2011-0031558)
programs through NRF funded by MSIP of Korea.

\bibliography{References}
\end{document}